\documentclass[%
 reprint,
superscriptaddress,
showpacs,preprintnumbers,
 amsmath,amssymb,
 aps,
pre,
floatfix,
]{revtex4-1}

\usepackage{graphicx}
\usepackage{color}
\usepackage{dcolumn}
\usepackage{bm}

\begin{document}


\title{Inferring Network Structure from Cascades}

\author{Sushrut Ghonge}
\email{sushrutghonge@gmail.com}
\affiliation{Department of Physics, Indian Institute of Technology Delhi, India}
\affiliation{
 Department of Physics, University of Notre Dame, USA}

\author{Dervis Can Vural}
\email{dvural@nd.edu}
\affiliation{
 Department of Physics, University of Notre Dame, USA}

\date{\today}

\begin{abstract}
Many physical, biological and social phenomena can be described by cascades taking place on a network. Often, the activity can be empirically observed, but not the underlying network of interactions. In this paper we offer three topological methods to infer the structure of any directed network given a set of cascade arrival times. Our formulas hold for a very general class of models where the activation probability of a  node is a generic function of its degree and the number of its active neighbors. We report high success rates for synthetic and real networks, for several different cascade models.

\end{abstract}

\maketitle

\section{Introduction}
Neural networks, ecosystems, epidemics, range expansions, gene-protein interactions, diffusion in evolutionary landscapes and many other interesting biological and social phenomena are naturally encoded by cascades on complex networks. Often we can observe when people adopt certain ideas, but we cannot see what social exchanges lead to it. We can observe when species go extinct, but do not know why \cite{RefWorks:doc:58042590e4b00630408e6b98}.
We see when the same content appears in several websites and blogs over time, but since we do not know who copied from whom, we cannot tell who follows who
\cite{RefWorks:doc:5804269be4b02929e7d62388}. A sequence of neural firings can be observed by flourescent imaging, but it is not trivial to infer neural connectivity \cite{brain}.

Establishing network structure empirically is tedious. Ideally, to determine the presence of an edge between a node pair, one must perturb one while measuring the response of the other, making sure the rest is unchanged. Proxies such as correlation coefficients can be used, but these yield unreliable results (e.g. \cite{berry2014deciphering}). Furthermore, proxies depend on specific models, e.g. the presence of an edge could just as well imply a lack of correlation. 

The problem of topological inference has been previously addressed as a convex optimization problem, and only specific cases have been solved
\cite{RefWorks:doc:58042590e4b00630408e6b98, RefWorks:doc:580425bae4b03fafe55c98c7,gomez2012submodular}.
Others have considered inferring topology when each cascade affects only few nodes, and only when few several such cascades take place simultaneously
\cite{RefWorks:doc:58042646e4b02929e7d6237b}.

This work concerns with a very general class of cascade models where the probability that a node activates depends on the degree of the node and the states of the neighboring nodes. In this class of models the activation of every node is permanent till the cascade ends. We  present three very generally applicable methods to determine network structure from time-of-activation data.  We then evaluate our success for 3 real networks, synthetic random networks and for 5 different kinds of cascade models. For one of the models we evaluate success for a full range of model parameters.

We cite an incomplete list of the systems and models for which our methods are applicable \cite{RefWorks:doc:580425e4e4b03fafe55c98cc, RefWorks:doc:5804269be4b02929e7d62388, RefWorks:doc:58042674e4b020ca61fde893,RefWorks:doc:5804265ce4b020ca61fde88e,RefWorks:doc:58042646e4b02929e7d6237b,gomez2012submodular,RefWorks:doc:58042603e4b03fafe55c98d3,RefWorks:doc:580425bae4b03fafe55c98c7,RefWorks:doc:58042590e4b00630408e6b98,RefWorks:doc:580b99b3e4b05487ad4f53da,RefWorks:doc:580b9979e4b05487ad4f53d1,RefWorks:doc:580b9965e4b026ac20c827a4,RefWorks:doc:580b990ee4b0995d7330f76c,RefWorks:doc:580b98ece4b0995d7330f769,RefWorks:doc:580b988de4b0995d7330f764,RefWorks:doc:580b988de4b0995d7330f764,gruhl2004information,RefWorks:doc:58042674e4b020ca61fde893,RefWorks:doc:58042603e4b03fafe55c98d3,vural2014aging}. In some of these models, nodes activate when a critical fraction of their providers (in-neighbors) activate \cite{vural2014aging,RefWorks:doc:580425e4e4b03fafe55c98cc} . In others, nodes do not deterministically activate when the number of active providers meet a threshold; instead their probability of activation jumps to a different value \cite{gleeson2008cascades}. In several other models, every active node linearly adds to the activation probability of their common neighbor. In general, a node can respond to its neighbors arbitrarily. The problem of inferring network structure finds applications in many diverse areas such as biochemistry and bioinformatics \cite{RefWorks:doc:580b98ece4b0995d7330f769,RefWorks:doc:580b9965e4b026ac20c827a4,RefWorks:doc:580b9979e4b05487ad4f53d1,RefWorks:doc:580b99b3e4b05487ad4f53da}, political science \cite{RefWorks:doc:58042674e4b020ca61fde893}, social networks, blogs \cite{RefWorks:doc:580b988de4b0995d7330f764, RefWorks:doc:580425bae4b03fafe55c98c7,RefWorks:doc:5804269be4b02929e7d62388}, sociology \cite{RefWorks:doc:58042590e4b00630408e6b98,RefWorks:doc:580425e4e4b03fafe55c98cc} and modelling aging \cite{vural2014aging}

\section{Diffusion model}
We consider the general model where the probability that a node activates is an arbitrary function $f(m/k)$ of the ratio of the number of active providers $m$ and its indegree $k$. 

We denote the fraction of nodes that activate at time $t$ by $D(t)$ and the fraction of nodes active at $t$ by $Q(t)$, so that $Q(t)=\sum_{\tau=1}^{t} D(\tau)$. 

For both the forward solution and the topological inversion, the probability that $m$ out of $k$ providers of a node have activated after a time $t$, can be approximated as
$B(m,k,Q(t))=\binom{k}{m}(Q(t))^{m}(1-Q(t))^{k-m}$, where $Q(t)$ is the fraction of nodes active at $t$. $Q(t)$ is to be determined recursively. 
 
For a node with $k$ providers, the probability $D(t)$ of activating is the sum over all possible number of activated providers of the product of the probability of that number of providers being active and the value of $f$ at that number. Since $\Gamma(k)$ is the fraction of nodes with indegree $k$, for a random node with unknown indegree,
\begin{align}\label{forward}
D(t)= \sum_{k} \Gamma(k) \sum_{m=0}^{k} B(m,k,Q(t-1))f(m/k)
\end{align}
Since all nodes are inactive at $t=0$, $D(1)=f(0)$. For $t \geq 2$, $D(t)$ is obtained in terms of $Q(t-1)= \sum_{\tau=1}^{t-1}D(i)$ from equation (\ref{forward}) which is easily iterated. This recursive equation was studied in detail in \cite{gleeson2008cascades}. 

Interestingly, knowing the forward dynamics gives little hint about the inverse problem of obtaining network topology, given node activation times.  This is an ill posed problem: generally speaking, two different networks (even those with different $\Gamma$) can have similar mean field behavior. Thus, the methods we develop will be probabilistic, i.e. we will output the network structure which is \emph{most likely} according to the method used.

\section{Topological Inversion}
We assume that an unknown network undergoes cascades numerous times and that we are given the times when each node activates in each cascade. Throughout, we will short-handedly denote a directional connection from $i$ to $j$, and that lack of, as $\overrightarrow{ij}$ and $\stackrel{\not\to}{ij}$ respectively.

Bayes theorem is frequently used in inverse problems related to networks. It has been successfully applied in several problems where network properties need to be inferred 
\cite{RefWorks:doc:580b99b3e4b05487ad4f53da,RefWorks:doc:580b9979e4b05487ad4f53d1,
RefWorks:doc:580b9965e4b026ac20c827a4,
RefWorks:doc:580b990ee4b0995d7330f76c,
RefWorks:doc:580b98ece4b0995d7330f769}. Bayesian methods have also been used to infer Bayesian networks \cite{friedman2003being}.

Let $N$ and $E$ be the network size and edge number. In the absence of any information, the probability that $\overrightarrow{ij}$ for randomly chosen nodes $i$ and $j$ is the fractional edge density (ratio of edges to number of possible edges).
$$P(\overrightarrow{ij} \mid \Omega)=\frac{E}{N(N-1)}\equiv\omega$$
were $\Omega$ denotes absence of information.

The Bayes theorem is used to update our probabilities when new information arrives. For events A and B, it states that
$P(A|B) = P(A|\Omega)P(B|A)/P(B)$. In the present problem, when we get the data from the first experiment giving us the time when $i$ and $j$ activate (let $E_1$ denote this event), the theorem gives us
\begin{align}
P_{1;i\to j}&=\omega P(E_1 |\overrightarrow{ij})/P(E_1)\\
P(E_1)&=\omega P(E_1|\overrightarrow{ij})+(1-\omega)P(E_1|\stackrel{\not \to}{ij})\nonumber
\end{align}
We update our probabilities iteratively. As more cascades happen we get more pairs of times for the activation of $i$ and $j$. 
\begin{align}
P_{n;i\to j}=P_{n-1;i\to j}P(E_n|\overrightarrow{ij})/P(E_n)
\end{align}
where, 
\begin{align}
P(E_n)&=P_{n-1;i\to j}P(E_n|\overrightarrow{ij})
+P_{n-1;i \not\to j}P(E_n|\stackrel{\not\to}{ij}),\nonumber\\
P_{n;i\to j}&=P(\overrightarrow{ij}| t_{i;1},t_{j;1},t_{i;2},t_{j;2},\cdots t_{i;n},t_{j;n}),\nonumber\\
P_{n;i \not\to j}&=1-P_{n;i\to j}\nonumber
\end{align}

After all experiments are completed, we will get a probability corresponding to each ordered pair of nodes $P(\overrightarrow{ij} \mid \text{all data})$, and choose $E$ edges with the highest probabilities and infer that they must be true edges.

We must now find how the probability that one node activates at $t_{1}$ and the other at $t_{2}$ is affected by the presence of a directed edge between the two nodes. To do so, we offer two methods: {\bf (M1)} obtaining it theoretically and {\bf (M2)} obtaining it semiempirically from a surrogate network with similar statistical properties. We can also infer networks heuristically without using Bayes Theorem {\bf (M3)}. The latter method has the advantage that it does not require the degree distribution of the network, but has less overall success and requires more experiments. We find that it is possible to use {\bf (M3)} to obtain the degree distribution when its success is above $80\%$ and then use this as an input for {\bf (M1)} or {\bf (M2)} which give far superior outcomes.

We evaluate the success of our three methods in Fig. 2 in detail for a particular forward model. We evaluate our success in Table 1 for other forward models. In all cases more number of experiments give higher overall accuracy. We supplement this letter with the working code that implements these methods. Further details of our three methods are outlined below.


\begin{figure*}
\includegraphics[width=17cm]{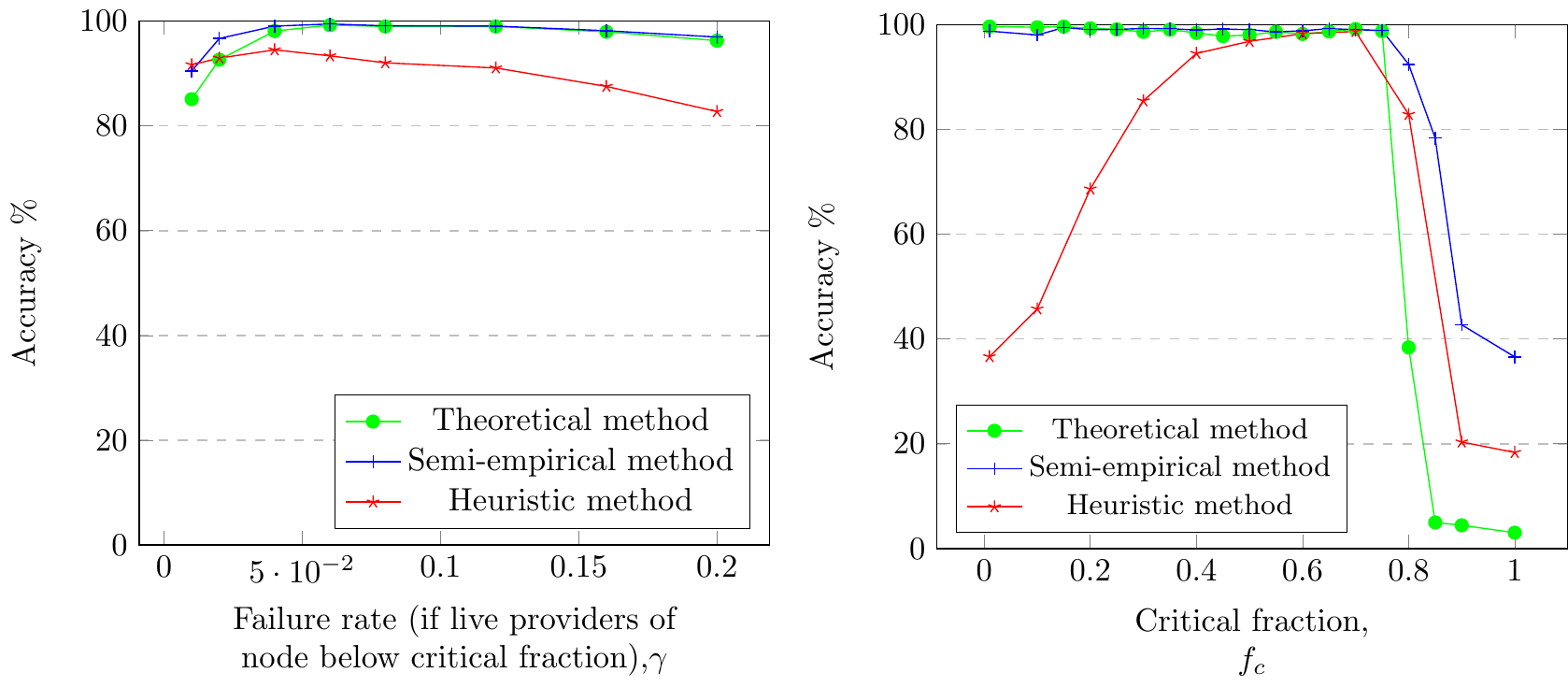}
\includegraphics[width=17cm]{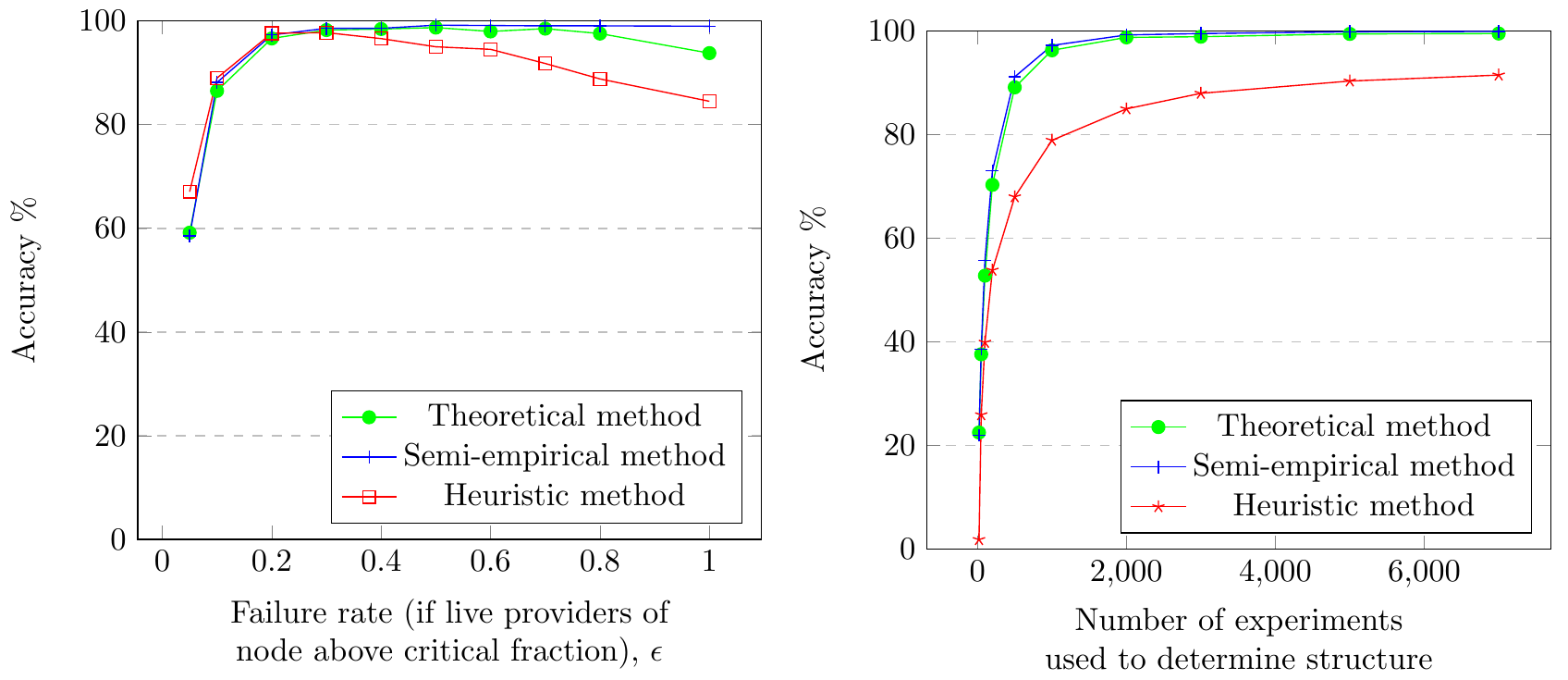}
\caption{{\bf Accuracy versus Model Parameters, for a specific model.} We evaluate our success rate using the three reported methods here, for a threshold model $f(m/k)$ that is equal to $\gamma$ for $m/k<f_c$ and $\epsilon$ when $m/k>f_c$. We sweep the parameter space and plot success rate as a function $\gamma$, $\epsilon$, $f_c$, (while keeping the other two constant at $\gamma=0.04$, $\epsilon=0.6$ $f_{c}=0.4$). For all runs the edges and network size are $E=1484$ and $N=200$. Number of experiments(cascades) is 2000 for Semiempirical and Theoretical methods except in the bottom right plot where we plot accuracy vs. number of experiments. For Heuristic method, the number of experiments is appropriately chosen according to the bottom right plot to give high accuracy. We evaluate other models $f(m/k)$ in Table 1}
\end{figure*}
\begin{figure*}
\includegraphics[width=17.5cm]{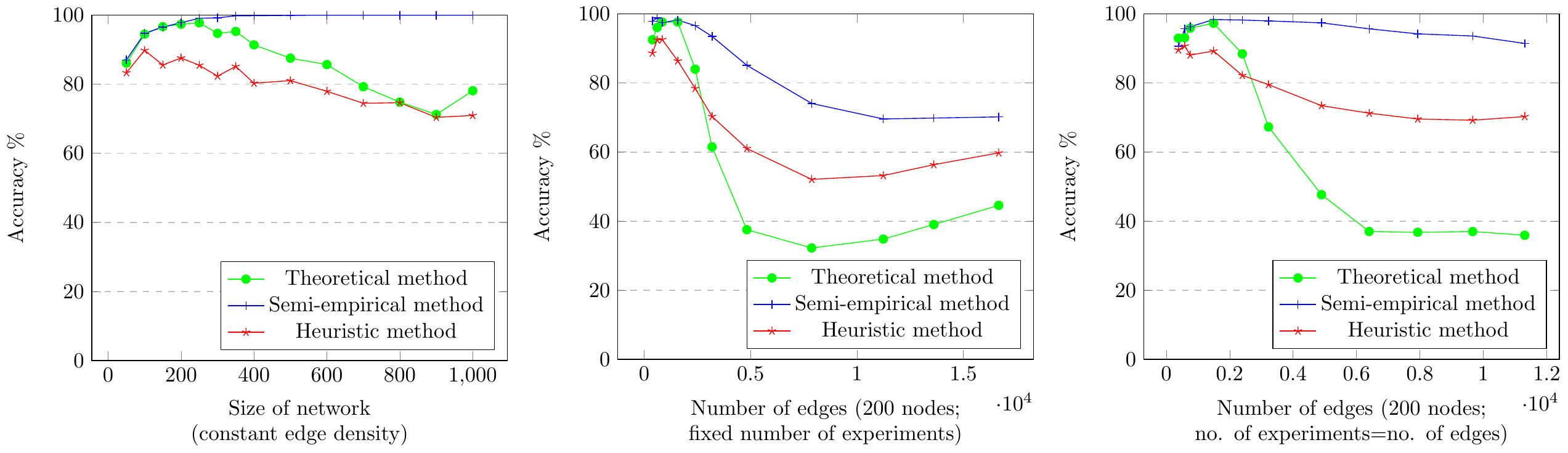}
\caption{{\bf Scalability of our methods, for a specific model.} We evaluate our success rate for networks of various sizes and densities. The threshold model with $f_c=0.4$, $\gamma=0.04$ and $\epsilon=0.6$ is used for cascades. Number of experiments(cascades) is 1600 in the first plot, is equal to the number of edges in the second plot and is equal to 0.4 times the number of possible edges (i.e. $0.4N(N-1)$) in the last plot. Edge density if fixed at 4\% in the last plot.}
\end{figure*}


{\bf (1) Theoretical Method}.
Here we theoretically derive an approximation for $P(t_{i},t_{j} \mid \overrightarrow{ij})$ and $P(t_{i},t_{j} \mid \stackrel{\not\to}{ij})$. Let $\overrightarrow{ij}$, and $j$ have an indegree $k$. At a time step when $i$ is inactive, $j$ has a total of $k-1$ providers which could possibly have activated. We assume all of them to be equivalent (i.e. equally likely to have activated). The probability that $m$ of those providers have activated at the given time will be a binomial distribution.

After $i$ has activated, there are still $k-1$ providers to choose from but there is an extra node which has activated. So the probability that $j$ is active is given by
\begin{align}
Q_{j}(t)=\sum_{k} \Gamma(k) \sum_{m=0}^{k-1} B(m,k-1,Q(t-1))h(t)
\end{align}
Where,
\begin{align}
h(t)=
\begin{cases}
f(m/k) & t\leq t_{i}\\
f((m+1)/k) & t>t_{i}
\end{cases}
\end{align}

To find $P(t_{i},t_{j} \mid \overrightarrow{ij})$, we need the probability that $j$ activates exactly at $t_{j}$. This is equal to the difference of the probabilities that $j$ is not active at $t_{j}-1$ and  the probability that it is active at $t_{j}$. We multiply this by the probability that $i$ activates at $t_{i}$.

$$P(t_{i},t_{j} \mid \overrightarrow{ij})= P(t_{j} \mid \overrightarrow{ij} \cap t_{i})P(t_{i})$$
$$P(t_{i},t_{j} \mid \overrightarrow{ij})= D(t_{i})\big[Q_{j}(t_{j})-Q_{j}(t_{j}-1)\big]$$
Note that since the activation time of each node in each cascade is known, $D(t)$ and $Q(t)$ can be obtained by simply counting the number of activations at that time.

In a large network, the activation of two arbitrarily chosen nodes at two different times are approximately independent. So $P(t_{1},t_{2})\approx D(t_{1})D(t_{2})$ for two times $t_{1}$ and $t_{2}$. This observation is used to obtain $P(t_{i},t_{j} \mid \stackrel{\not \to}{ij}$ which is required in the Bayes' theorem (3), as follows-
$D(t_{i})D(t_{j})=\omega P(t_{i},t_{j}|\overrightarrow{ij})
+(1-\omega )P(t_{i},t_{j}|\stackrel{\not\to}{ij})$\\

{\bf (2) Semiempirical Method}
This is a simple method in which we construct another (surrogate) network with similar statistical properties. Now we can do as many experiments on this network to ``measure'' $P(t_i,t_j \mid \overrightarrow{ij})$ for all times. Then we use the values of this function in (1) and (2) to get a probability for every entry in the connectivity matrix to be a true edge. A network with the same indegree distribution as that of the unknown network can be easily constructed by starting with an empty network and adding random edges to every node one by one until the exact degree distribution is reached.

{\bf (3) Heuristic Method.} When the degree distribution is not known, the edges can be considered to be pairs of random variables. Some methods have been developed to infer network structures by finding joint information or correlation between these variables \cite{RefWorks:doc:5804265ce4b020ca61fde88e}. 
Here we exploit the observation that if a node activates at some time, it is quite likely that one or more of its providers activated just before it. We find how often one node activates right after another, and choose the edges between nodes with highest number of such consecutive activations.

\section{Evaluation}
To test the accuracy of our methods we simulated various models on known synthetic and real networks and used the activation time of nodes from the simulations as if experimental data. We then compared our inferred networks to the actual ones.

\begin{table}
\begin{tabular}{|c || c | c | c |}
\hline
$g(m/k)$ &Theoretical&Semiempirical&Heuristic\\
\hline
$m/k$&98.78&99.12&95.05\\
$(m/k)^{2}$& 95.89 & 99.12&95.96\\
$1-(1-m/k)^{2}$&99.66&99.80&90.36\\
$1-exp(-m/k)$&97.71&97.57&88.54\\
$1-exp(-3m/k)$&99.80&99.73&81.67\\
\hline
\end{tabular}
\caption{Accuracy(\%) of inversion for some general models of the kind  $f(m/k)=0.04+0.96g(m/k)$, for N=200, E=1563, obtained from 2000 cascades on a random network.}
\end{table}

\begin{table}
\begin{tabular}{|c || c | c | c || c | c | c |}
\hline
{} &\multicolumn{3}{c|}{Threshold Model} &\multicolumn{3}{c|}{$g(m/k)=m/k$}\\
\hline
Expt. &Theo.&SE&Heur.&Theo.&SE&Heur.\\
\hline
$20$&59.34&64.84&0&48.90&63.74&0\\
$50$&78.02&80.22&0&70.33&79.67&45.05\\
$100$&84.62&84.62&73.63&78.57&87.36&68.68\\
$200$&86.81&85.16&76.92&89.01&91.76&84.62\\
$500$&88.46&88.46&85.71&90.11&91.21&86.26\\
\hline
\end{tabular}
\caption{Accuracy(\%) of inference for Gagnon and Macrae prison network: N=67, E=182, for when {\bf left:} f(m/k) is a step function (threshold model) with lower value $\gamma=0.04$, higher value $\epsilon=0.6$ and threshold point $f_{c}=0.4$ and {\bf right:} $f(m/k)=0.04+0.96m/k$}.
\end{table}

\begin{table}
\begin{tabular}{|c || c | c || c | c|}
\hline
{} & \multicolumn{2}{c|}{Advice,E=480} & \multicolumn{2}{c|}{Discussion,E=565}\\
Experiments & Theoretical &SE& Theoretical &SE\\
\hline
$25$&56.67&62.08&55.40&59.29 \\
$50$&73.12&78.33&72.92&74.69 \\
$100$&82.92&83.12&82.83&83.36 \\
$200$&87.92&88.12&89.03&87.61 \\
$500$&92.71&71.45&92.92&83.62 \\
\hline
\end{tabular}
\caption{Accuracy(\%) of inference for physician networks (N=246): Threshold model with $\epsilon=0.6$, $f_{c}=0.4$ $\gamma$=4\%}
\end{table}

As a first example, we inverted a generalized version of \cite{RefWorks:doc:580425e4e4b03fafe55c98cc} such that $f(m/k)=\gamma$ if $m/k<f_{c}$ and $\epsilon$ if $m/k \geq f_{c}$. In other words, a node changes its activation probability if more than a critical number of providers activate. We varied all model parameters for this example and plotted our accuracy in Fig.1. We also demonstrate the scalability of our methods for this example in Fig. 2, where we vary the size of the network and the number of edges.
\begin{table}
\begin{tabular}{|c || c | c | c|}
\hline
Experiments & Theoretical &SE &Heuristic\\
\hline
$100$&64.74&79.49&63.78\\
$200$&65.38&84.61&69.23\\
$500$&69.87&85.90&73.72 \\
$1000$&72.75&85.90&76.28\\
\hline
\end{tabular}
\caption{Accuracy(\%) of inference for Zachary's Karate club network: N=34, E=156, $\gamma$=4\%, $\epsilon=0.6$, $f_{c}=0.4$}.
\end{table}

In all plots and tables, we do not report accuracy as defined by the fraction of correctly identified connectivity matrix elements, but fraction of correctly identified edges. For example, in a network of $100$ nodes and $100$ edges we must decide whether $\sim10^4$ entries of the connectivity matrix is a $0$ or $1$. If we identify 10 false edges (and hence, also not identify 10 true edges), we report our accuracy rate as $90/100=90\%$ instead of $9980/10^4=99.8\%$.
In addition to the threshold model we also evaluate others models (without varying all possible parameters of these models). Our success rates are reported in Table I.

We tried to infer friendships between inmates of the Gagnon and Macrae prison using synthetic data. The network consists of 67 prisoners(nodes) which have 182 friendships (edges) \cite{macrae1960direct}. Success rates are reported in Table II.
We have also used our methods on some undirected graphs such as the Zachary's Karate club network \cite{zachary1977information}. It has 34 members of a karate club (nodes) and have 156 friendships (edges) between members. The experiments simulated resemble studying the spread of an opinions and practices among friends. $\gamma$ is included to represent opinion formation due factors other than friends, success rates are reported in Table IV.

Several physicians were surveyed in \cite{burt1987social} and \cite{coleman1957diffusion} to study how information about a new medicine spreads among physicians that do friendly discussions or take professional advice. This was later modeled as a network problem in \cite{valente1995network}, and effects of marketing were studied in \cite{van2001medical}. In our simulations, $\gamma$ simulates the effect of marketing and the jump at $f_{c}$ assumes that a physician starts prescribing a medicine with an increased probability $\epsilon$ if their colleagues prescribe it. The results of inferring physicians' relationships with their colleagues using synthetic data of cascades (i.e. medicine prescriptions) is given in Table III.

See Supplemental Material at
\color{blue}
\href{http://link.aps.org/supplemental/10.1103/PhysRevE.96.012319}{(PhysRevE.96.012319)}
\color{black} for computer programs of all of our inversion methods and instructions for using them.  

\section{Limitations}
We conclude our study by discussing our limitations. Our methods do not produce accurate results when the critical fraction is so high that most nodes activate not due to interactions, but randomly. Since in this case, the structure of the network plays little role in the cascade dynamics, it becomes difficult to extract the structure. We also observe that the theoretical method does not work well for very dense networks (Fig. 2). This is because our (approximate) formulas depend only on the indegree distribution $\Gamma (k)$. However, in dense networks, higher order, conditional indegree distributions (such as the probability that a degree $k$ node has a degree $m$ connection $\Gamma (k,m)$) plays an important role. The semiempirical method works best for random networks and its success is slightly lower for other kinds of networks. This is because we match only the indegree distribution of the surrogate network and the outdegree distribution may not be well matched for other kinds of networks. This method is essentially a binary classification of individual edges, but we can also calculate conditional probabilities of trees in the network using Bayes theorem. Relying only on binary classification leads to poor accuracies at higher link densities where higher order structures like trees and cycles play a major role in cascade propagation. Another limitation can be seen in Table IV, where as the number of experiments increases, the accuracy may decrease. This is a common and well-known issue with naive Bayesian classifiers \cite{rish2001empirical}. Lastly, the binomial approximation in (\ref{forward}) works less successfully in networks for which the providers have different likelihood of activating. Nevertheless, our success with heterogeneous networks (cf. Tables II-IV) show that this inaccuracy is not very crucial.
\bibliography{export}
\end{document}